# Evaluation of Pervasive Games: Recruitment of Qualified Participants through Preparatory Game Phases

Vlasios Kasapakis, Damianos Gavalas and Thomas Chatzidimitris

University of the Aegean. Department of Cultural Technology and Communication, Mytilene
{v.kasapakis,d.gavalas,tchatz}@aegean.com

**Abstract.** In this paper we present the evaluation process for *Barbarossa*, a pervasive role playing game. *Barbarossa* involves an invitational (preparatory) and a main execution phase. The former is freely available though Google Play store and may be played anytime/ anywhere. The latter defines three inter-dependent player roles acted by players who need to collaborate in a treasure hunting game. The eligibility of players for participating in the main game phase is restricted among those ranked relatively high in the invitational phase. Herein, we investigate the impact of the invitational game mode on the players overall game experience. The main hypothesis tested is that game awareness (gained from participating in a preliminary game phase) may serve as a means for recruiting the most suitable subjects for user trials on pervasive game research prototypes.

## 1 Introduction

As pervasive game prototypes proliferate and gamers' awareness on this emerging gaming genre consolidates, research focus is increasingly diverted towards understanding the human factors which are mostly influential to the overall players' experience. Along this line, several challenges surface with respect to pervasive games evaluation and several evaluation methods have been proposed and tested in order to support designers' comprehension on the aspects that impact the overall quality of experience in pervasive games [1, 2]. Game evaluation trials serve as a valuable instrument for measuring the enjoyment and immersion perceived by players; however, they typically involve a lengthy and expensive multi-phase process (preparatory activities, subjects' recruitment, trials orchestration and monitoring, execution and compilation of surveys, etc). The evaluation process is even more demanding when considering pervasive game trials, wherein trials are executed in the physical space, far from a supervised laboratory environment. As a result, user trials should be well prepared and carefully orchestrated to ensure their flawless execution.

The focal objective of evaluations trials is to receive unbiased feedback from neutral, 'external' subjects about the usability, playability and experience perceived throughout the game sessions. The recording of neutral views expressed by the evaluators may indicate technical flaws or even suggest essential script, usability or technical improvements. Hence, the recruitment of qualified participants is regarded as one of the most critical and challenging aspects of user evaluation trials [2]. Another crucial, yet commonly overlooked, aspect of pervasive games evaluation relates with players invitations. Montola argues that the participation awareness level of a player during game sessions commonly shifts from the 'unaware' to the 'aware' state. The level of awareness may be influenced by the invitation method employed to recruit players. Furthermore, addressing invitations to prospective players (i.e. those that belong to the game's potential target group) may allow them to gain game experiences while increasing participation awareness [3].

The pervasive game prototypes evaluated in the past have utilized a variety of invitation methods for recruiting participants like e-mails [4, 5], personal contacts [6], announcements/advertisements [4,7], recruitment of colleagues/organization employees [8], Jones and Marsden tabulated a list of advantages and disadvantages inherent in the above subject recruitment methods [9].

Notably, none of the above referenced evaluated pervasive games utilized invitation methods that enhance players' game awareness. That is, experiences are missing in assessing evaluation methods which actually introduce the players into the game; even more so, no methods have been proposed to allow game designers monitoring and 'screening' the players and provide them the means to select the most eligible players to participate into the evaluation process.

The main hypothesis investigated in this article is that a preparatory game phase (acting as a 'qualification round') would designate the players mostly interested in participating in the main game phase; hence, these players would be the most appropriate evaluators as they could be regarded as representative sample of the game's potential target group. The above hypothesis has been validated in the evaluation of *Barbarossa*. Along this line we opted to implement a freely available invitational game mode in the Android application market - Google Play , enabling players worldwide to participate into the game, thereby ensuring openness and neutrality in the recruitment process. Also by strongly linking the invitational phase game to the overall game scenario, the players have been seamlessly introduced into the game and gained a better understanding of its concept and goals.

## 2   Game Scenario & Implementation

*Barbarossa* [10] is a two-phase trans-reality role playing game. The first game phase is available from the Google Play app store under the title "The Conqueror"[1]. In the first phase scenario the Barbarossa pirate brothers Aruj, Khzir and Ilyas (known for their pirate raids throughout the Aegean sea during 1600-1650 A.D.), following a battle against the Knights of St. John outside the castle of Mytilene and assisted by some traitors within the castle walls, conquered the city. In this phase the players act as Knights of the St. John who try to free the conquered city. Acting so, the players use a custom Android application which utilizes Google Maps and a turn-based role-playing game which allows them to complete and create quests located into the surrounding area of Mytilene. Upon completing quests the players gain experience points that indicate their commitment and attribution to the game.

---

[1] https://play.google.com/store/apps/details?id=zarc.crash.conqueror&hl=en

The players ranked higher (in experience points) in the first phase are invited to participate in the second game phase called "The Interplay"; in the latter, Mytilene is freed and the Knights rush in the castle to catch Aruj, Ilyas and Khzir. At the crypts of the castle the three Pirates knowing that the Knights are looking for them, hide all their treasures to a treasure chest. Aruj and Ilyas lock the chest with one combination lock each, while Khzir takes it and ridding his horse leaves the crypts to hide it. In a while, the Knights arrive at the crypts and manage to catch Aruj as a prisoner. Ilyas, though, manages to flee and a Knight chases after him.

In order to complete the second game phase, three players should cooperate to unlock the treasure chest hid by Khzir somewhere in the city. It is noted that the "The Conqueror" application detects the distance of the players from the Mytilene center (located on Lesvos island, Greece) while playing, grouping players into two separate experience rankings categories, the Insiders who play in Mytilene ($\leq 3.5$ km from the city center) and the Outlanders who play away from the city. One of the Outlanders and another of the Insiders as well as a guest (selected by the Insider player) are invited into the second phase based on their total experience points, collected in the first phase. The three players utilize custom Android applications integrating a variety of technologies, including QR-Code scanning, environment sound level recording, augmented reality, location-based gaming, the Google Directions service, sensor devices (SunSPOTs), etc. The players aim at completing their assigned missions (based on separate, yet, supplementary scenarios) to locate the locked chest and the two lock combinations and unlock the chest. Full implementation details of *Barbarossa* may be found in the official website of the game at www.BarbarossaRPG.com.

## 3  Evaluation Method & Results

So far, questionnaires, interviews and log data (i.e. data capturing the mobility and interaction activity of players throughout the game sessions) have been the methods most commonly employed in pervasive games evaluation. The same practice has been followed in *Barbarossa*. We have conducted user evaluation trials using all the three abovementioned evaluation methods; log data have been a critical element in the evaluation process in order to cross-check them (when available) against player answers (as compiled by questionnaires and interviews) and extract more safe and reliable conclusions. Below we describe the evaluation process in full detail. For the questionnaires we used linkert scale and yes/no questions.

The evaluation process of the game commenced in October $29^{th}$, 2013 by releasing the invitational game mode though Google Play as well as a website wherein the players could check their rankings. We provided players sufficient time (21 days) to play the first game phase; thereafter we started contacting the highest ranked players among the Outlanders and the Insiders in order to form the 3-player teams required to proceed to the second game phase. We have invited one team at a time, a practice that enhanced competition among Phase I players who wished to participate in the second game phase.

Prior to proceeding to the second game phase we have asked all members of the 3-player teams that qualified from Phase I (i.e. those acting as Treasure Hunters and Knights) to complete a questionnaire about their experiences in the first phase. Then, we introduced the players into the second game phase and allowed them a week to play the game session and collaboratively locate and unlock the treasure chest. Having completed the second phase, we have invited all players to complete an additional questionnaire tailored to the scenario they pursued in the second game phase. Finally, each player has been interviewed about her overall game experience. In parallel, we have collected log data (e.g. total completed and created quests, game session duration, distance travelled and speed) about the players game actions throughout the game sessions.

Currently (July 2014) *Barbarossa* features more than 1500 downloads and a average rating of 4.04/5 in Google Play. Furthermore, 874 players registered in *Barbarossa*; according to our log data, 262 among them performed at least one in game action, such as undertaking a quest.

Before investigating the impact of the invitational phase on the overall game experience, the usability aspects of the invitational game mode had to be evaluated to ensure that no serious flaws prevented players state transition towards game awareness. Table 1 presents the responses of evaluators with respect to their gaming background. Almost all players stated that they are regular video game players, while only two first phase participants have had previous experience with games similar to *Barbarossa*.

**Table 1.** Demographic questions.

| Do you play video games regularly? | | |
|---|---|---|
| Yes (Y) | 15 | 75.00% |
| No (N) | 5 | 25.00% |
| **Have you played a game similar to *Barbarossa* in the past?** | | |
| Yes (Y) | 2 | 10.00% |
| No (N) | 18 | 90.00% |

We have also addressed several questions to participants to understand their perception of game usability aspects of the first phase game mode (see Table 2).

**Table 2.** Usability questions.

| Overall, the game system performed well with no serious errors or flaws | | |
|---|---|---|
| Strongly agree | 9 | 45.00% |
| Agree | 10 | 50.00% |
| Neutral | 1 | 5.00% |
| Disagree | 0 | 0.00% |
| Strongly disagree | 0 | 0.00% |
| **It was easy to find, download, install and start the game** | | |
| Strongly agree | 17 | 85.00% |
| Agree | 3 | 15.00% |
| Neutral | 0 | 0.00% |
| Disagree | 0 | 0.00% |
| Strongly disagree | 0 | 0.00% |
| **It was easy to learn and recall how to perform basic actions in the game** | | |
| Strongly agree | 15 | 75.00% |
| Agree | 5 | 25.00% |
| Neutral | 0 | 0.00% |
| Disagree | 0 | 0.00% |
| Strongly disagree | 0 | 0.00% |

Our evaluation results revealed that the invitational game mode performed well without any serious flaws that could affect players' experience. The above finding is backed by the only 6 error reports submitted by players to the Google Play Developer Console. Besides, the results indicate easy access to the invitational game mode. Finally, all players responded positively with respect to the games learnability (i.e. their ability to recall how to perform basic game actions).

After the completion of the second game phase we invited the players to complete questionnaires about the scenario they pursued in that phase followed by an interview. In order to assess the impact of the invitational mode to the players' comprehension of the game goals we asked the players to express their perception about the clarity of the game goals. As illustrated in Table 3, the players (even the Knights who did not met their co-players in person and played far from the game stage location) were aware of the game goals and also felt responsible to complete their mission to support the success of their team.

**Table 3.** Overall game play experience & social/multiplayer aspects questions.

| **The game goal was comprehensible and unambiguous** | | | | |
|---|---|---|---|---|
| | **Knight** | | **Treasure Hunter** | |
| Strongly agree | 9 | 90.00% | 10 | 100.00% |
| Agree | 1 | 10.00% | 0 | 0.00% |
| Neutral | 0 | 0.00% | 0 | 0.00% |
| Disagree | 0 | 0.00% | 0 | 0.00% |
| Strongly disagree | 0 | 0.00% | 0 | 0.00% |
| **I felt responsible to complete my mission for my team to succeed** | | | | |
| | **Knight** | | **Treasure Hunter** | |
| Strongly agree | 7 | 70.00% | 9 | 90.00% |
| Agree | 2 | 20.00% | 1 | 10.00% |
| Neutral | 1 | 10.00% | 0 | 0.00% |
| Disagree | 0 | 0.00% | 0 | 0.00% |
| Strongly disagree | 0 | 0.00% | 0 | 0.00% |

The most interesting questions investigating the impact of the first game phase have been asked during the interview, as we opted to allow participants to freely and fully express their views. In Table 4 we present the interview results concerning players' opinion on the utility of the invitational phase. Compiled participant answers are presented as a percentage of positive and negative answers. Answers are grouped by the player assigned roles into the second phase. The Pirate players have not been inquired about the first game phase as they have not participated to it (they have been invited into the second phase by their Treasure Hunters friends).

**Table 4.** Interview questions about the invitation phase.

| Question | Was the first phase of the game useful in order to understand the whole game concept? | Did the first phase of the game eased the second phase of the game completion? | Having played the first phase of the game, have you developed interest on how the game would progress? |
|---|---|---|---|
| **Treasure Hunters** | Yes (100%) | Yes (100%) | Yes (100%) |
| **Knights** | Yes (100%) | Yes (100%) | Yes (100%) |

The results clearly indicate that all players admitted the invitational game mode impact in raising their awareness on the overall game concept as well as assisting the completion of the second phase scenario. Finally the invitational game mode triggered players' interest on the game's progression, thereby increasing their keenness to participate in Phase II.

## 4 Conclusion

Pervasive game prototype developers traditionally relied on emails, personal contacts and announcements/advertising to invite participants and perform user trials. In *Barbarossa*, we utilized an invitational game mode to recruit qualified participants for the user evaluation trials. The recruited participants provided valuable feedback and represented both players located worldwide (as the first phase of *Barbarossa* has been freely available online) and also located in the area where the game has been actually staged (Mytilene).

The evaluation results confirmed that the execution of a preparatory game mode, when applicable, can help developers to recruit highly qualified participants, truly enthusiastic to playing the game. Further, invitational game modes may serve as a useful instrument for developers to train evaluation participants on any technological equipment used in the game and also enhance their awareness on the overall game goal, scenario and gameplay.